\documentclass[prb,reprint,superscriptaddress]{revtex4-1}
\usepackage{graphicx}
\usepackage{bm}
\usepackage{color}

\begin{document}

\title{Effect of grain-boundary diffusion process on the geometry of the grain microstructure of Nd$-$Fe$-$B nanocrystalline magnets}

\author{Ivan Titov}\email[Electronic address: ]{ivan.titov@uni.lu}
\author{Massimiliano Barbieri}
\author{Philipp Bender}
\author{Inma Peral}
\affiliation{Physics and Materials Science Research Unit, University of Luxembourg, 162A~Avenue de la Fa\"iencerie, L-1511~Luxembourg, Grand Duchy of Luxembourg}
\author{\\Joachim Kohlbrecher}
\author{Kotaro Saito}
\affiliation{Paul Scherrer Institute, CH-5232~Villigen PSI, Switzerland}
\author{Vitaliy Pipich}
\affiliation{Forschungszentrum J\"ulich GmbH, J\"ulich Centre for Neutron Science (JCNS) at Heinz Maier-Leibnitz Zentrum (MLZ), \\ Lichtenbergstra{\ss}e~1, D-Garching~85748, Germany}
\author{Masao Yano}
\affiliation{Advanced Material Engineering Division, Toyota Motor Corporation, Susono~410-1193, Japan}
\author{Andreas Michels}\email[Electronic address: ]{andreas.michels@uni.lu}
\affiliation{Physics and Materials Science Research Unit, University of Luxembourg, 162A~Avenue de la Fa\"iencerie, L-1511~Luxembourg, Grand Duchy of Luxembourg}

\date{\today}

\begin{abstract}
Hot-deformed anisotropic Nd$-$Fe$-$B nanocrystalline magnets have been subjected to the grain-boundary diffusion process (GBDP) using a $\mathrm{Pr}_{70}\mathrm{Cu}_{30}$ eutectic alloy. The resulting grain microstructure, consisting of shape-anisotropic Nd$-$Fe$-$B nanocrystals surrounded by a Pr$-$Cu-rich intergranular grain-boundary phase, has been investigated using unpolarized small-angle neutron scattering (SANS) and very small-angle neutron scattering (VSANS). The neutron data have been analyzed using the generalized Guinier-Porod model and by computing model-independently the distance distribution function. We find that the GBDP results in a change of the geometry of the scattering particles:~In the small-$q$ regime the scattering from the as-prepared sample exhibits a slope of about $2$, which is characteristic for the scattering from two-dimensional platelet-shaped objects, while the GBDP sample manifests a slope of about $1$, which is the scattering signature of one-dimensional elongated objects. The evolution of the Porod exponent indicates the smoothing of the grain surfaces due to the GBDP, which is accompanied by an increase of the coercivity.
\end{abstract}


\maketitle\

\section{Introduction}

Nd$-$Fe$-$B based nanocrystalline permanent magnets are of potential interest for electronic devices, motors, and wind turbines due to their preeminent magnetic properties such as high coercivity and magnetic energy product~\cite{coehoorn88,Davies1989,gutfleisch2002,gutfleisch2011,liu2009,bance2014}. In this context, the effect of the grain-boundary diffusion process (GBDP) on the bulk magnetic microstructure of hot-deformed Nd$-$Fe$-$B based nanomagnets is currently extensively investigated (e.g., Ref.~\cite{liu2013,SepehriAmin2013,liu2014,SepehriAmin2015,sala2018}). In the GBDP~\cite{SepehriAmin20101124}, the Nd$-$Fe$-$B magnet is exposed at elevated temperatures to a fine powder or a vapor containing high-magnetic-anisotropy-inducing heavy-rare-earth elements such as Tb or Dy, which then diffuse (preferentially along liquid grain boundaries) into the bulk of the material, in this way locally increasing the coercivity.

The goal of the present work is to obtain microscopic information about the nanoscale structure of Pr$-$Cu infiltrated $\mathrm{Nd}_2\mathrm{Fe}_{14}\mathrm{B}$ nanocrystalline alloys by means of unpolarized small-angle neutron scattering (SANS) and very small-angle neutron scattering (VSANS). The SANS technique (see Ref.~\cite{rmp2019} for a recent review) is ideally suited for monitoring the changing grain-boundary chemistry due to the GBDP~\cite{yano2014no1,yano2014no2,saito2015,perigo2016no2}, since it provides---in contrast to electron-microscopy methods---statistically-averaged bulk information about both the structural and magnetic correlations on a mesoscopic length scale ($\sim 1-1000 \, \mathrm{nm}$). This method has previously been applied to study the structures of magnetic nanoparticles~\cite{disch2012,guenther2014,bender2015,bender2018jpcc,bender2018prb,oberdick2018,krycka2019}, soft magnetic nanocomposites~\cite{suzuki2007,herr08pss}, proton domains~\cite{michels06a,aswal08nim,noda2016}, magnetic steels~\cite{bischof07,bergner2013,Pareja:15,shu2018}, or Heusler-type alloys~\cite{bhatti2012,runov2006,michelsheusler2019}. Here, we aim to correlate the changes in the grain microstructure due to the GBDP with the coercivity.

\section{Experimental}

\begin{figure}[tb!]
\centering
\resizebox{0.85\columnwidth}{!}{\includegraphics{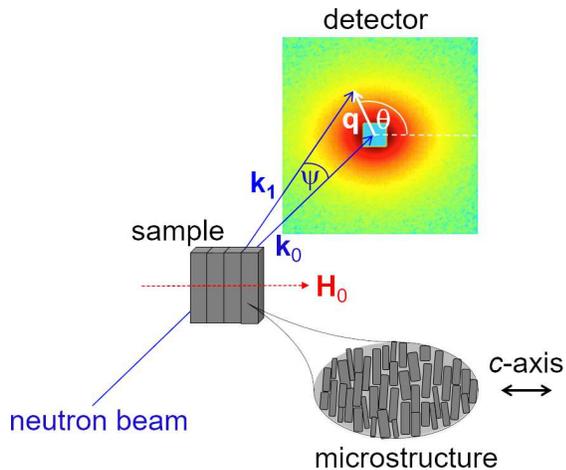}}
\caption{Schematic drawing of the SANS setup. The scattering vector $\mathbf{q}$ is defined as the difference between the wave vectors of the scattered and incident neutrons, i.e., $\mathbf{q} = \mathbf{k}_1 - \mathbf{k}_0$; its magnitude $q = |\mathbf{q}| = (4\pi/\lambda) \sin(\psi/2)$ depends on the mean wavelength $\lambda$ of the neutrons and on the scattering angle $\psi$. The applied-field direction $\mathbf{H}_0$ is parallel to the $\mathbf{e}_z$-direction of a Cartesian laboratory coordinate system and perpendicular to the incident neutron beam ($\mathbf{k}_0 \parallel \mathbf{e}_x \perp \mathbf{H}_0$). In the small-angle approximation, the component of $\mathbf{q}$ along $\mathbf{k}_0$ is neglected, i.e., $\mathbf{q} \cong \{0, q_y, q_z\} = q\{0, \sin\theta, \cos\theta\}$, where the angle $\theta$ specifies the orientation of $\mathbf{q}$ on the two-dimensional detector. The pressing direction ($c$-axis) of the samples is horizontal ($\parallel \mathbf{H}_0$).}
\label{fig1}
\end{figure}

We have studied a Pr$-$Cu-doped series of hot-deformed Nd$-$Fe$-$B nanocrystalline magnets. The specimens were prepared by the melt-spinning technique. The resulting melt-spun ribbons were crushed into powders of a few hundred micrometer sizes and then subsequently sintered at $1048 \, \mathrm{K}$ under a pressure of $400 \, \mathrm{MPa}$. The hot-pressing procedure results in the formation of shape anisotropic $\mathrm{Nd}_2\mathrm{Fe}_{14}\mathrm{B}$ grains which are stacked along the nominal $c$-axis (see the sketch of the microstructure in Fig.~\ref{fig1}). The Pr$-$Cu-doped series contains an undoped reference sample and two samples which are doped with 20~wt.\% and 40~wt.\% of Pr$_{70}$Cu$_{30}$. In the following these samples are, respectively, referred to as PrCu0, PrCu20, and PrCu40. More details on the sample preparation can be found in Refs.~\cite{liu2013,SepehriAmin2013,liu2014,SepehriAmin2015}. Magnetization data were taken on a $14 \, \mathrm{T}$ vibrating sample magnetometer.

The neutron experiment has been carried out at $300 \, \mathrm{K}$ at the instrument SANS-I at the Paul Scherrer Institute (PSI), Switzerland, using unpolarized neutrons with a mean wavelength of $\lambda = 5.0 \, \mathrm{\AA}$ and $\Delta \lambda / \lambda = 10 \, \mathrm{\%}$ (FWHM)~\cite{kohlbrecher2000,niketic2015}. The external magnetic field $\mathbf{H}_0$ was applied perpendicular to the wave vector $\mathbf{k}_0$ of the incoming neutron beam ($\mathbf{k}_0 \perp \mathbf{H}_0$); see Fig.~\ref{fig1} for a sketch of the neutron setup. This corresponds to the geometry where $\mathbf{H}_0$ is parallel to the nominal $c$-axis (pressing direction) of the textured samples. Neutron data were corrected for background scattering (empty sample holder), transmission, and detector efficiency using the GRASP software package~\cite{graspurl}. The measured neutron transmission of all samples was larger than $80-90 \, \mathrm{\%}$ at all fields investigated. The SANS setup at PSI allowed us to access the following range of momentum transfers: $0.02 \,\mathrm{nm}^{-1} \lesssim q \lesssim 1.0 \,\mathrm{nm}^{-1}$. To significantly reduce the minimum momentum-transfer value to $q_{\mathrm{min}} \cong 0.003 \,\mathrm{nm}^{-1}$, additional unpolarized runs were carried out at the very small-angle neutron scattering (VSANS) instrument KWS-3 operated by the J\"ulich Centre for Neutron Science (JCNS) at the Heinz Maier-Leibnitz Zentrum (MLZ), Garching, Germany~\cite{kws3}. In the VSANS experiments we used a mean wavelength of $\lambda = 12.8 \, \mathrm{\AA}$ [$\Delta \lambda / \lambda = 16 \, \mathrm{\%}$ (FWHM)] and a sample-to-detector distance of $9.8 \, \mathrm{m}$.

\section{SANS cross section and correlation function}

For $\mathbf{k}_0 \perp \mathbf{H}_0$, the unpolarized elastic nuclear and magnetic differential SANS cross section $d \Sigma / d \Omega$ at momentum-transfer vector $\mathbf{q}$ reads~\cite{rmp2019}:
\begin{eqnarray}
\label{sigmatotperp}
\frac{d \Sigma}{d \Omega}(\mathbf{q}) = \frac{8 \pi^3}{V} b_H^2 \left( b_H^{-2} |\widetilde{N}|^2 + |\widetilde{M}_x|^2 + |\widetilde{M}_y|^2 \cos^2\theta \right. \nonumber \\ \left. + |\widetilde{M}_z|^2 \sin^2\theta - (\widetilde{M}_y \widetilde{M}_z^{\ast} + \widetilde{M}_y^{\ast} \widetilde{M}_z) \sin\theta \cos\theta \right) ,
\end{eqnarray}
where $V$ denotes the scattering volume, $b_H = 2.91 \times 10^{8} \, \mathrm{A^{-1} m^{-1}}$, $\widetilde{N}(\mathbf{q})$ represents the nuclear scattering amplitude, $\widetilde{\mathbf{M}}(\mathbf{q}) = \{ \widetilde{M}_x(\mathbf{q}), \widetilde{M}_y(\mathbf{q}), \widetilde{M}_z(\mathbf{q}) \}$ is the Fourier transform of the magnetization $\mathbf{M}(\mathbf{r}) = \{ M_x(\mathbf{r}), M_y(\mathbf{r}), M_z(\mathbf{r}) \}$, the asterisk ``$\,^{\ast}\,$'' marks the complex-conjugated quantity, and $\theta$ is the angle between $\mathbf{H}_0$ and the scattering vector $\mathbf{q}$.

In the literature on small-angle scattering there exist many approaches to analyzing anisotropic scattering patterns (see, e.g., Refs.~\cite{summerfield1983,mildner1983,reynolds1984,hammouda1986a,hammouda1986b,saraf1989,svetogorsky1990,gu2016,gu2018} and references therein). Here, the SANS data were analyzed in terms of the generalized Guinier-Porod model, which has been developed by Hammouda~\cite{hammouda2010gp} in order to describe the $2 \pi$ azimuthally-averaged scattering from nonspherical objects (such as rods or lamellae). The model is purely empirical and essentially decomposes the $I(q) = \frac{d \Sigma}{d \Omega}(q)$ curve into a Guinier region for $q \leq q_1$ and into a Porod region for $q \geq q_1$. Both parts of the scattering curve are then joined by demanding the continuity of the Guinier and Porod laws (and of their derivatives) at $q_1$; more specifically~\cite{hammouda2010gp},
\begin{eqnarray}
\label{gpmodeleq1}
I(q) &=& \frac{G}{q^s} \exp\left( - \frac{q^2 R_G^2}{3-s} \right) \hspace{0.25cm} \mathrm{for} \hspace{0.25cm} q \leq q_1 \\
I(q) &=& \frac{D}{q^n} \hspace{0.25cm} \mathrm{for} \hspace{0.25cm} q \geq q_1 ,
\end{eqnarray}
where the scaling factors $G$ and $D$, the Guinier radius $R_G$, the dimensionality factor $s$, and the Porod power-law exponent $n$ are taken as independent parameters. From the continuity of the Guinier and Porod functions and their derivatives it follows that:
\begin{eqnarray}
\label{gpmodeleq2a}
q_1 &=& \frac{1}{R_G} \left[ \frac{(n-s)(3-s)}{2} \right]^{1/2} \\
D &=& G q_1^{n-s} \exp\left( - \frac{q_1^2 R_G^2}{3-s} \right) ,
\label{gpmodeleq2}
\end{eqnarray} 
where $n>s$ and $s<3$ must be satisfied. Note that $q_1$ is not a fitting parameter, but internally computed via Eq.~(\ref{gpmodeleq2a}).

The generalized Guinier-Porod model is commonly applied to orientationally-averaged microstructures. In the Supplemental Material to this paper~\cite{smtitov2019} we demonstrate that it can also be used to describe the small-angle scattering from oriented particles:~the $2\pi$-averaged one-dimensional SANS cross sections from oriented cylinders and ellipsoids are similar to the corresponding SANS cross sections of the randomly oriented ensembles. Moreover, it is shown that synthetic data on oriented cylinders and ellipsoids with a distribution of sizes can be described by the generalized Guinier-Porod model.

In addition to the above analysis using the generalized Guinier-Porod model, we have model-independently calculated the distance distribution function~\cite{bender2017}
\begin{equation}
\label{pvonrfunc}
p(r) = r^2 \int_0^{\infty} \frac{d \Sigma}{d \Omega}(q) j_0(q r) q^2 dq ,
\end{equation}
where $j_0(z) = \sin(z)/z$ denotes the zeroth-order spherical Bessel function. This provides information on the characteristics (e.g., size and shape) of the scattering objects~\cite{svergun03,glatter2006}, and on the presence of interparticle correlations~\cite{glatter1996,glatter2011}.

\section{Results and discussion}

\begin{figure}[tb!]
\centering
\includegraphics[width=0.90\columnwidth]{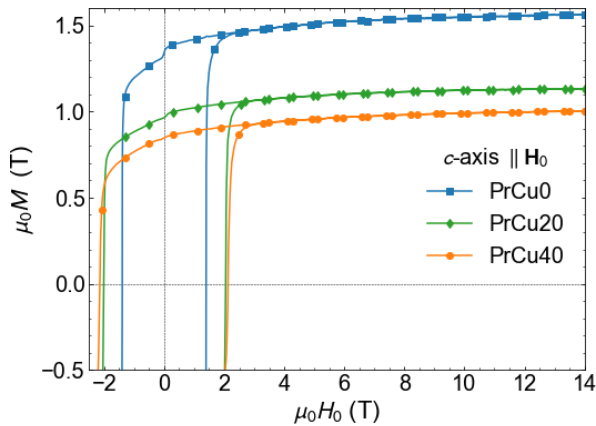}
\caption{Effect of Pr$-$Cu grain-boundary diffusion process on the room-temperature magnetization of hot-deformed Nd$-$Fe$-$B based nanocrystalline magnets. Applied-field direction is along the $c$-axis (pressing direction). For the clarity of presentation, the number of data points has been reduced.}
\label{fig2} 
\end{figure}

Figure~\ref{fig2} displays the effect of Pr$-$Cu infiltration on the room-temperature magnetic hysteresis of hot-deformed Nd$-$Fe$-$B. As expected and in agreement with data in the literature~\cite{liu2013,SepehriAmin2013,liu2014,SepehriAmin2015}, the coercivity $H_c$ increases significantly with increasing Pr$-$Cu content (from $\mu_0 H_c = 1.39 \, \mathrm{T}$ for PrCu0 to $\mu_0 H_c = 2.03 \, \mathrm{T}$ for PrCu20 to $\mu_0 H_c = 2.14 \, \mathrm{T}$ for PrCu40) at the cost of lowering the saturation magnetization and the remanence.

\begin{figure*}[tb!]
\centering
\includegraphics[width=1.90\columnwidth]{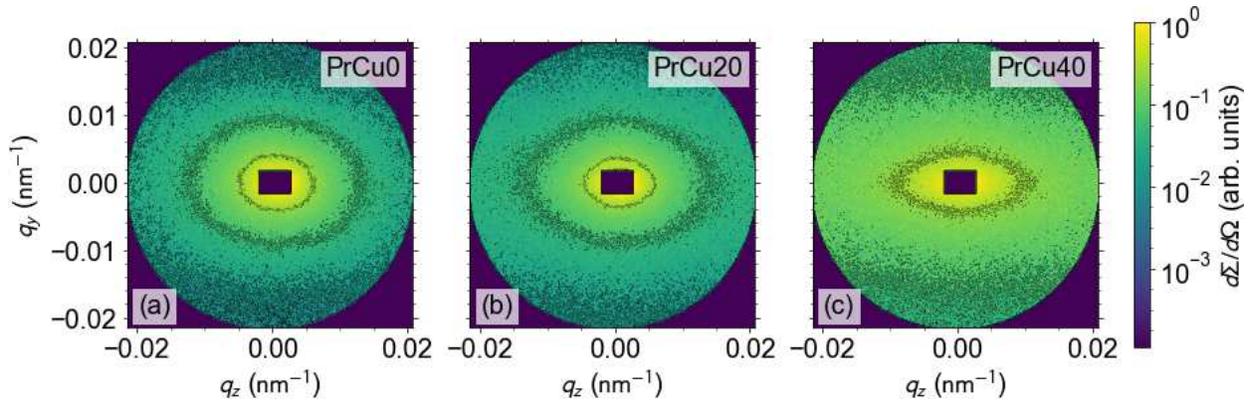}
\caption{Two-dimensional total nuclear and magnetic VSANS cross section $d \Sigma / d \Omega$ of Pr$-$Cu doped hot-deformed Nd$-$Fe$-$B nanocrystalline magnets (logarithmic color scale). Data in the remanent state are shown ($H_0 = 0$). Applied-field direction was along the horizontal direction, parallel to the $c$-axis (pressing direction).}
\label{fig3} 
\end{figure*}

\begin{figure*}[tb!]
\centering
\includegraphics[width=1.90\columnwidth]{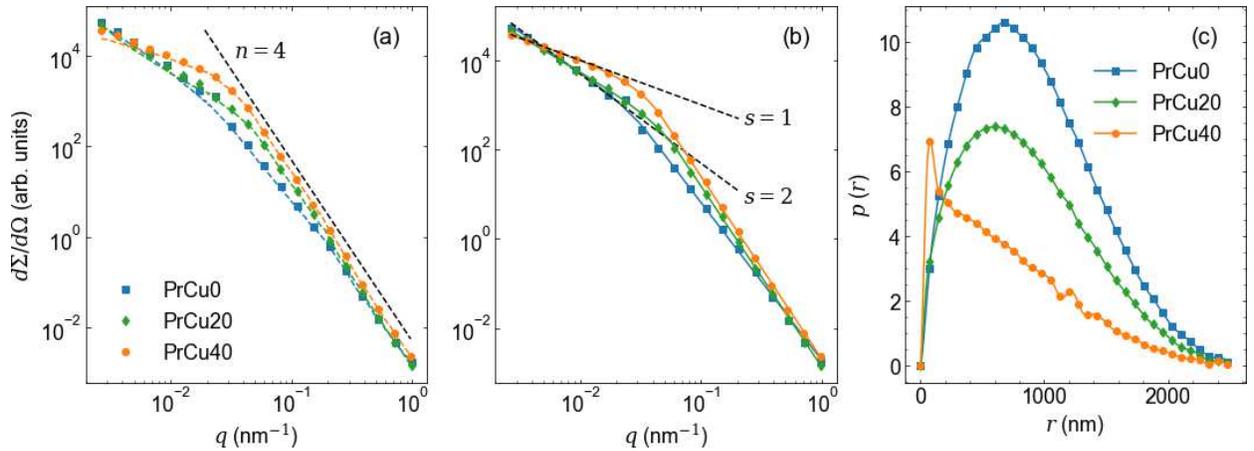}
\caption{(a)~Azimuthally-averaged total SANS cross sections $d \Sigma / d \Omega$ of Pr$-$Cu doped hot-deformed Nd$-$Fe$-$B nanocrystalline magnets (log-log scale). VSANS data from KWS-3 at MLZ and ``conventional'' SANS data from SANS-I at PSI have been merged. Solid markers:~Remanent-state data; dashed lines:~$\mu_0 H_0 = 2.2 \, \mathrm{T}$. (b)~Same as panel~(a), but for the remanent state. Solid lines: Fit to the generalized Guinier-Porod model [Eqs.~(\ref{gpmodeleq1})$-$(\ref{gpmodeleq2})]. Dotted lines: Asymptotic power-law behavior according to two-dimensional and one-dimensional scatterers in the generalized Guinier-Porod model. (c)~Distance distribution functions $p(r)$ [Eq.~(\ref{pvonrfunc})] of the remanent-state data. For the clarity of presentation, the number of data points has been reduced in (a)$-$(c).}
\label{fig4} 
\end{figure*}

The two-dimensional VSANS data are displayed in Fig.~\ref{fig3}, whereas the one-dimensional (over $2\pi$ azimuthally-averaged) data are shown in Fig.~\ref{fig4}. From the observation in Fig.~\ref{fig4}(a) that the scattering curves at the remanent state and at an applied field of $\mu_0 H_0 = 2.2 \, \mathrm{T}$ (dashed lines) are basically identical, it follows that $d \Sigma / d \Omega$ is dominated by the field-independent nuclear scattering, i.e., $d \Sigma / d \Omega \propto |\widetilde{N}|^2$. Regarding the two-dimensional patterns in Fig.~\ref{fig3} this then implies that the origin of the pronounced angular anisotropy is related to an anisotropic (oriented) grain microstructure, which in turn is in line with the results of electron-microscopy investigations~\cite{liu2013,SepehriAmin2013,liu2014,SepehriAmin2015,michelspra2017}. With increasing Pr$-$Cu content we see, in particular at the small momentum transfers $q$, that the two-dimensional $d \Sigma / d \Omega$ increase their horizontal and decrease their vertical elongation. This finding points towards a change of the geometry of the scattering objects due to the GBDP. While all the samples exhibit a similar power-law decay at the large $q$-values (essentially with a Porod exponent of $n \cong 4$, see below), there is a pronounced change in the slope of the scattering with doping in the small $q$-regime [compare Fig.~\ref{fig4}(a)]. Such changes in the slope of $d \Sigma / d \Omega$ within the small-$q$ Guinier or the intermediate-$q$ Guinier regime indicate a change in the dimensionality of the scattering objects~\cite{hammouda2010gp}.

In order to gain further information on the changes in the geometry of the scattering particles, we have analyzed the SANS cross sections in the remanent state using the generalized Guinier-Porod model [Eqs.~(\ref{gpmodeleq1})$-$(\ref{gpmodeleq2})]. The solid lines in Fig.~\ref{fig4}(b) are the result of this analysis and the fitting parameters are listed in Table~\ref{tab1}. The most remarkable observation is the change in the $s$-parameter, from $1.80$ for PrCu0 to $0.94$ for PrCu40. This suggests that the elements of the microstructure which give rise to $d \Sigma / d \Omega$ change their geometry (shape)~\cite{hammouda2010gp}: from two-dimensional platelet-shaped, corresponding to $s = 2$, to one-dimensional rod-like ($s = 1$). The change in $s$ is also observed when averages of $d \Sigma / d \Omega$ along the horizontal (easy $c$-axis) and perpendicular (hard axis) directions are taken. The effective Guinier radius varies between $R_G = 27.7 \, \mathrm{nm}$ (PrCu0), $R_G = 19.6 \, \mathrm{nm}$ (PrCu20), and $R_G = 37.1 \, \mathrm{nm}$ (PrCu40). This quantity is a measure for the smallest dimension of the scatterers, which sensitively depends on the particle-size distribution and on the particle shape, e.g., $R_G = R/\sqrt{2}$ for the cross section of a randomly oriented cylinder with radius $R$, while $R_G = T/\sqrt{12}$ for the cross section of a randomly oriented lamella with thickness $T$~\cite{feigin,hammouda2010gp}. Therefore, in view of this sensitivity, we cannot relate the variation of $R_G$ found from the fit to the actual changes in the real-space dimensions of the particles (see also the discussion in~\cite{smtitov2019}). But the $R_G$-values are within the expected size range, as seen by electron microscopy~\cite{liu2013,SepehriAmin2013,liu2014,SepehriAmin2015,michelspra2017}. The Porod exponent $n$ increases from $n = 3.53$ for PrCu0 to $n \cong 4.1$ for the infiltrated samples. This indicates a smoothing of the surface of the particles as a result of the Pr$-$Cu infiltration~\cite{hammouda2010gp}.

\begin{table}[tb!]
\caption{\label{tab1} Results of a weighted nonlinear least-squares fit using the generalized Guinier-Porod model [Eqs.~(\ref{gpmodeleq1})$-$(\ref{gpmodeleq2})]. The values for the scaling parameters $G$ and $D$ are not listed; $q_1$ is not a fit parameter, but computed according to Eq.~(\ref{gpmodeleq2a}).}
\begin{ruledtabular}
\begin{tabular}{lcccc}
Sample & $s$ & $R_G$~($\mathrm{nm}$) & $n$ & $q_1$~($\mathrm{nm}^{-1}$) \\
\hline
\hline
PrCu0 & $1.80 \pm 0.02$ & $27.7 \pm 0.9$ & $3.53 \pm 0.01$ & $0.0369$ \\
PrCu20 & $1.60 \pm 0.01$ & $19.6 \pm 0.3$ & $4.06 \pm 0.01$ & $0.0669$ \\
PrCu40 & $0.94 \pm 0.02$ & $37.1 \pm 0.5$ & $4.09 \pm 0.01$ & $0.0485$ \\
\end{tabular}
\end{ruledtabular}
\end{table}

The results for the distance distribution function $p(r)$ in Fig.~\ref{fig4}(c) are consistent with the numerical fit analysis using the generalized Guinier-Porod model. The PrCu0 and the PrCu20 samples both exhibit a $p(r)$ which is typical for globular, but slightly anisotropic particles, whereas the $p(r)$ of the PrCu40 sample shows a maximum at small $r \cong 100 \, \mathrm{nm}$ followed by a long tail at the larger $r$, suggesting that the scattering originates from shape-anisotropic elongated objects (compare Fig.~5 in the review by \textcite{svergun03}). The maximum of $p(r)$ at the small distances of the PrCu40 specimen corresponds to the shortest dimension of the particles. This behavior is consistent with the previously observed dimensional crossover ($s \cong 2 \rightarrow s \cong 1$) using the Guinier-Porod model.

\section{Conclusion}

Unpolarized small-angle neutron scattering (SANS) and very small-angle neutron scattering (VSANS) were used to monitor the effect of the grain-boundary diffusion process (GBDP) on the mesoscopic grain microstructure of Nd$-$Fe$-$B nanocrystalline magnets. The SANS data, which are predominantly of nuclear origin, reveal a pronounced effect of the GBDP, resulting in a change of the slope of $d \Sigma / d \Omega$ with increasing doping at small momentum transfers. Analysis of the scattering data in terms of the generalized Guinier-Porod model suggests that the GBDP results in a dimensional crossover of the geometry of the scattering objects: The SANS data from the as-prepared specimen are characteristic for two-dimensional platelet-shaped objects, while the doped samples manifest the signature of one-dimensional rod-like objects. This assessment is further supported by an independent model-free analysis in terms of the distance distribution. Moreover, based on the evolution of the Porod exponent, we find an indication for the smoothing of the grain surfaces due to the GBDP, which goes along with an increase of the coercivity.

\section*{Acknowledgment}

Philipp~Bender and Andreas~Michels acknowledge financial support from the National Research Fund of Luxembourg (CORE~SANS4NCC~grant). Kotaro~Saito has received funding from the European Union’s Horizon 2020 research and innovation programme under the Marie Sk{\l}odowska-Curie grant agreement No.~701647. This paper is based on results obtained from the future pioneering program ``Development of magnetic material technology for high-efficiency motors'' commissioned by the New Energy and Industrial Technology Development Organization (NEDO). The neutron experiments were performed at the Swiss spallation neutron source SINQ, Paul Scherrer Institute, Villigen, Switzerland and at the Heinz Maier-Leibnitz Zentrum, Garching, Germany.

\bibliographystyle{apsrev4-2}

%

\end{document}